\documentstyle [12pt,epsf]{article}

\input{epsf}
\begin{document}
\begin{titlepage}
\begin{center}

{\Large \bf The thermal coupling constant and the  
gap equation in the $\lambda\varphi^{4}_{D}$ model}\\
%\vspace{.3in}
{\large\em G.N.J.A\~na\~nos\footnotemark[1], A.P.C.Malbouisson\footnotemark[2]
and N.F.Svaiter\footnotemark[3]}\\
Centro Brasileiro de Pesquisas F\'{\i}sicas-CBPF\\ Rua Dr.Xavier
 Sigaud 150, Rio de Janeiro, RJ 22290-180 Brazil\\
\end{center}
\vspace{-0.4in}
\subsection*{Abstract}
By the concurrent use of two different  
resummation methods, the composite operator formalism  
and the Dyson-Schwinger equation, we re-examinate the behavior at finite 
temperature of the $O(N)$-symmetric $\lambda\varphi^{4}$ model 
in a generic D-dimensional Euclidean space.   
In the cases $D=3$ and $D=4$, an analysis of the thermal behavior 
of the renormalized squared mass and coupling constant are done for all
temperatures. It results that the thermal renormalized squared mass 
is positive and increases monotonically with the temperature. The 
behavior of the thermal coupling constant is quite different in odd or 
even dimensional space. In $D=3$,  
the thermal coupling constant decreases up to a 
minimum value diferent from zero and then grows up monotonically as the 
temperature increases. In the case $D=4$, it is found that the thermal renormalized coupling constant  tends in the 
high temperature limit to a constant asymptotic value. Also for 
general D-dimensional Euclidean space,
we are able to obtain a formula for the critical temperature 
of the second order phase transition. This formula agrees with 
previous known values at $D=3$ and $D=4$. 
\footnotetext[1]{e-mail:gino@lafex.cbpf.br}
\footnotetext[2]{e-mail:adolfo@lafex.cbpf.br} \footnotetext[3]{e-mail:nfuxsvai@lafex.cbpf.br}

\end{titlepage}

\newpage

\baselineskip .37in
\section {Introduction}

Considerable progress has been done during the last two decades in our 
understanding of finite temperature field theory. 
A clear account on the 
foundations of this presently well established branch of 
theoretical physics is done in Refs.\cite{kapu}. One of the outstanding 
questions for the subject is the possible 
existence of a deconfinement phase 
transition in QCD. It is expected that at sufficiently high temperatures 
quarks and gluons are not bounded inside hadrons, but have a behavior somehow 
analogous to the plasma phase, the so called quark-gluon plasma \cite{pol}.
In connection to this expected phenomenon 
it naturally arises questions concerning the 
nature of the transition from the low temperature colorless hadron gas 
to the high temperature quark-gluon plasma \cite{kala}. 
In particular, if in fact it is a phase transition, 
the question of its order is still 
open. In pure Yang-Mills theory it is a point of consensus that there 
is a phase transition, but if quarks are included, the answer seems 
to be more difficult. Also if the theory involves quarks and gluons, another 
phase transition may occur and its order depends on the number 
of quark flavours: the chiral symmetry phase transition. The chiral 
symmetry is spontaneously broken at zero temperature and it is 
restored at some finite temperature. For two massless flavours it is 
a second order one, while for three flavours it is a first order 
phase transition. For more than three flavours no clear conclusion 
has been obtained about its order, nor even if there is a chiral phase transition. Another important point to be noticed in the the case of the 
chiral symmetry phase transition in QCD, is that the order parameter 
is composite 
in the fields and perturbative methods fail. Non-perturbative methods are mandatories.

In any case, ``real" QCD is 
very difficult to deal with and no definite theoretical 
results for many questions 
have been obtained up to the present moment. So, simpler 
models still remain very useful ``laboratories" to try to get 
some insights about what happens in the real world, particularly if 
long range, non-perturbative effects are involved, as 
it should be the case for both phase transitions mentioned above. 
For instance,  
it is well known that the 
low energy dynamics of QCD may be quite well described by the 
$O(4)$ model, since there is an isomorphism between the $O(4)$ and the 
$SU(2)XSU(2)$ symmetry groups, the latter being the group of two flavors of 
massless quarks \cite{kap} (in the $O(N)$ model when the 
vacuum of the model exhibits spontaneous symmetry breaking it is 
known as the linear $\sigma$ model).

In a previous paper two of the authors derived 
expressions for the renormalized thermal 
mass and coupling constant in the $\lambda\varphi^{4}$ model 
in a D-dimensional
flat spacetime at the one-loop approximation \cite{adolfo}. 
The main results
are that the thermal squared mass increases with the 
temperature, while the 
thermal coupling constant decreases with the temperature. These 
authors conjectured that
in $D=3$ it could exist a temperature above which the renormalized coupling 
constant becomes negative. In this regime the system would develop a 
first order phase transition. As we will see latter, in the framework of the vector N-component model at a non-perturbative level, the answer seems to 
be negative. 
In other words, in this context, there is no 
temperature at which the coupling 
constant changes of sign, and no first order phase transition 
induced by the thermal renormalized coupling constant seems 
to be possible in $D=3$. 
On the other side, in a slightly different context, another result 
that deserves our attention was obtained by Ananos and Svaiter \cite{an}. 
These authors studying the $(\lambda\varphi^{4}+\sigma\varphi^{6})_3$ model at 
finite temperature up the 
two-loop approximation, exhibit the existence of the tricritical 
phenomenon when both, the thermal renormalized squared mass and quartic 
coupling constant become zero for some temperature. 

The purpose of this 
paper is to get an extension of 
the above mentioned result \cite{adolfo} to all orders of 
perturbation theory. 
This may be done considering the $O(N)$-symmetric model in the 
framework of the large-N expansion at the leading order in $\frac{1}{N}$, and 
using resummation methods: the composite operator formalism (CJT) \cite{cornwall} and also the Dyson-Schwinger equation, both adapted to finite temperature. It is well known that these resummation methods 
can solve the problem of the  breakdown of perturbation theory in 
some massless
field theories at finite temperature. For instance, 
in the N-component scalar $\lambda\varphi^{4}$ model it is 
possible, for large N, to sum a class of Feynman diagrams, the ring diagrams, 
by the use of the recurrent Dyson-Schwinger equation. 
This allows to solve at the leading order in $\frac{1}{N}$, the 
problem of infrared divergences \cite{dolan}.  
An alternative method that takes into account 
leading and subleading contributions from multiloops diagrams was 
developed by Cornwall, Jackiw and Toumboullis (CJT formalism) \cite{cornwall}.
In this approach one considers a generalization of the effective action 
$\Gamma(\varphi)$ which depends not only on the expectation value of 
the field, but also on the expectation value of the time ordered product 
of two fields. In this formalism we naturally sum a large class 
of diagrams, and the gap equation is easily obtained using a variational 
technique. This formalism has been further developed by 
Pettini and more recently by Amelino-Camelia and Pi \cite{amelino}.
It may also be mentioned that recently discussions on the pure 
scalar model have been done by many authors \cite{gin}. Of 
special interest are Ref.\cite{reu} and Ref.\cite{novo}. In this last article 
Eylal et al. \cite{novo} 
discussed the phase structure of the $O(N)$ model in a 
generic D-dimensional 
euclidean space, and also the sign of the second coefficient of the 
large-N renormalization group $\beta$ function.
It may be noticed that Kessler and Neuberger \cite{kes} using lattice regularization obtained quite surprisingly 
that for $D=3$ the sign of this cofficient is arbitrary. 
The authors in Ref.\cite{novo} raised the possibility
that this ambiguity could be an artifact of the large N-expansion.  

In the next sections we apply the above mentioned resummation methods to 
get non-perturbative results for the renormalized thermal 
squared mass 
and coupling constant for the vector massive 
$\lambda\varphi^{4}$ model defined 
in a generic D-dimensional Euclidean space. The outline of the paper 
is the following. In section II we 
briefly discus the CJT formalism. In section III the thermal gap equation 
is obtained. In section IV using the gap equation we 
sum all the daisy and super-daisy diagrams to obtain the thermal 
renormalized coupling constant. Conclusions are given in section V. 
In this paper we use $\frac{h}{2\pi}=c=k_{B}=1$.

\section{The Cornwall, Jackiw and Tomboulis (CJT) formalism}\

In this section we will briefly discuss the effective action formalism 
for composite operators extended to finite temperature in a D-dimensional 
Euclidean space \cite{livro}.
Let us consider the vacuum persistence amplitude $Z(J,K)$ in the 
presence of local and non-local  
sources $J(x)$ and $K(x,y)$ respectively, where $J(x)$ couples to $\Phi (x)$
and $K(x,y)$ to $\frac{1}{2}\Phi(x)\Phi(y)$. This object is a  
generalized generating functional of the Green's functions of the model i.e.,
\begin{equation}
Z(J,K)=\int D\Phi\exp\left\{-\left[I(\Phi)+
\int d^{D}x \left(\Phi(x)J(x)\right)+\frac{1}{2}\int d^{D}x\int d^{D}y
\left(\Phi(x)K(x,y)\Phi(y)\right)\right]\right\},
\label{cjt1}
\end{equation}
where $D\Phi$ is an appropriate functional measure.
Introducing the interaction Lagrange density $L_{int}$,
the classical Euclidean action $I(\Phi)$  
is given by
\begin{equation}
I(\Phi)= \int d^{D}x\int d^{D}y \left(\Phi(x) D_{0}^{-1}(x-y)\Phi(y)\right)+
\int d^{D}x L_{int},
\label{cjt2}
\end{equation}
where $D_{0}(x-y)$ is the free propagator i.e.,

\begin{equation}
D_{0}^{-1}(x-y)=-(\Box +m^2 ) \delta^{D}(x-y).
\label{cjt3}
\end{equation}
The generalized 
effective action $\Gamma(\varphi,G)$ is defined by a double 
Legendre transform of the generalized generating functional of the connected 
correlation functions $W(\varphi,G)=\ln Z(\varphi,G) $ (the Helmholtz 
free energy), 
\begin{equation}
\frac{\delta W(J,K)}{\delta J(x)}=\varphi(x),
\label{cjt4}
\end{equation}
\begin{equation}
\frac{\delta W(J,K)}{\delta K(x)}=\frac{1}{2}\left( \varphi(x) \varphi(y)+ G(x,y)
\right),
\label{cjt5}
\end{equation}
and 
\begin{equation}
\Gamma(\varphi,G)= W(J,K)-\int d^{D}x \left(\varphi(x)J(x)\right)-\frac{1}{2} \int d^{D}x
\int d^{D}y \left(\varphi(x)\varphi(y)+G(x,y)\right)K(x,y).
\label{cjt6}
\end{equation}
In the absence of sources, the generalized effective action $ \Gamma(\varphi,G) $
satisfies 
\begin{equation}
\frac{\delta \Gamma(\varphi,G)}{\delta \varphi(x)}=J(x)-\int d^{D}y 
\left(K(x,y)
 \varphi(y)\right)=0
\label{cjt7}
\end{equation}
and
\begin{equation}
\frac{\delta \Gamma(\varphi,G)}{\delta G(x,y)}=-\frac{1}{2} K(x,y)=0.
\label{cjt8}
\end{equation}
In eq.(\ref{cjt4}) and eq.(\ref{cjt5}), the quantity $\varphi(x)$ is the normalized 
vacuum expectation value of the field,
\begin{equation}
\varphi(x) = < 0 | \Phi(x)| 0>  
\end{equation}
and $G(x,y)$ is the two-point function. 
If we assume that translation invariance is not spontaneously broken we have 
$\varphi(x) \equiv \varphi$ and the propagator depends only on the
distance in Euclidean space, $G(x,y) \equiv G(x-y)$.
Consequently it is possible to define the generalized effective potential
$V(\varphi,G)$ as a straightforward generalization of the usual 
definition,
\begin{equation}
\Gamma(\varphi,G)= -V(\varphi,G) \int d^{D}x .
\label{cjt9}
\end{equation}
The stability conditions given by eqs. (\ref{cjt7}) and (\ref{cjt8}) become 
in terms of the effective potential,
\begin{equation}
\frac{\partial V(\varphi,G)}{\partial \varphi}=0 \;\;\;\;\;\;,\;\;\;\;\;\; 
\frac{\partial V(\varphi,G)}{\partial G(k)}=0,
\label{10}
\end{equation}
where $G(k)$ is the Fourier transform of $G(x,y)$
\begin{equation}
G(x,y)=G(x-y)=\frac{1}{(2 \pi)^{D}} \int d^{D}k e^{-ik(x-y)}G(k)=
\frac{1}{(2 \pi)^{D}} \int d^{D}k e^{-ik(x-y)}\frac{1}{k^{2}+M^{2}},
\label{11}
\end{equation}
and $M^{2}=m^{2}+\frac{\lambda}{2}\varphi^{2}$.

Let us consider the model described by the Euclidean Lagrange density in the form 
\begin{equation}
{\cal L}=\frac{1}{2} \partial_{\mu}\Phi \partial_{\mu} \Phi + \frac{1}{2}
m^2 \Phi^2 + \frac{\lambda}{4!} \Phi^4, 
\label{14}
\end{equation}
whose effective potential is given by 
\begin{eqnarray}
V(\varphi,M)= \frac{1}{2} m^2 \varphi^2 + 
\frac{\lambda}{4!} \varphi^4 + \frac{1}{2}
 \int \frac{d^D k}{(2\pi)^D} \ln (k^2+M^2)- \\ \nonumber
\frac{1}{2} ( M^2-m^2-
\frac{\lambda}{2}\varphi^2)G(x,x) +\frac{\lambda}{8} G(x,x) G(x,x).
\label{15}
\end{eqnarray}
Combining the above equation with the stationary requirements given by 
eq.(\ref{10}) we have 
\begin{equation}
\frac{\partial V(\varphi,M)}{\partial \varphi}=\varphi \left( m^2 + \frac{\lambda}{6}\varphi^2+\frac{\lambda}{2} G(x,x) \right)=0 \; ,
\label{16}
\end{equation}
\begin{equation}
\frac{\partial V(\varphi,M)}{\partial M^2}= -\frac{1}{2} 
\frac{\partial G(x,x)}{\partial M^2}\left( M^2-m^2-\frac{\lambda}{2} \varphi^2 -\frac{\lambda}{2} G(x,x) \right)=0.
\label{17}
\end{equation}
The effective potential is obtained by evaluating $V(\varphi,M)$ from  
eq.(\ref{17}). It is composed by the classical,
the one-loop and two-loop contributions, i.e.,
\begin{equation}
V(\varphi,M(\varphi))=V_{0}+V_{I}+V_{II},
\label{iso1}
\end{equation}
where the classical contribution is 
\begin{equation}
V_{0}(\varphi)=\frac{1}{2} m^2  \varphi^2 + 
\frac{\lambda }{4!} \varphi^4, 
\label{iso2}
\end{equation}
the one loop contribution is 
\begin{equation}
V_{I}(\varphi,M(\varphi))=\frac{1}{2}
 \int \frac{d^D k}{(2\pi)^D} \ln (k^2+M^2(\varphi)),
\label{iso3}
\end{equation}
and finally the two-loop contribution is 
\begin{equation}
V_{II}(\varphi,M(\varphi))=
-\frac{\lambda }{8}G(x,x)G(x,x).
\label{iso4}
\end{equation}
In the above expressions we have,
\begin{equation}
M^{2}(\varphi)=m ^{2}+\frac{\lambda}{2}\varphi^{2}+
\frac{\lambda}{2}G(x,x),
\label{iso5}
\end{equation}
and $G(x,x)$ is given by
\begin{equation}
G(x,x)=\int \frac{d^D k}{(2 \pi)^D} \frac{1}{k^2+M^2(\varphi)}.
\label{iso6}
\end{equation}
To study finite temperature effects there are two different formalisms, 
the real time formalism and the imaginary time (Matsubara) formalism. In the Matsubara formalism the 
Euclidean time $\tau$ is
restricted to the interval $ 0 \leq \tau \leq \beta=\frac{1}{T}$ 
and in the functional integral the field $\Phi(\tau,{\bf x})$ satisfies
periodic boundary conditions in Euclidean time,  
\begin{equation}
\Phi(0,{\bf x})=\Phi(\beta,{\bf x}).
\label{13}
\end{equation}
All the Feynman rules are
the same as in the zero temperature case, except that the momentum-space 
integral over the zero-th component is replaced by a sum over discrete 
frequencies. For boson fields we have to perform the replacement
\begin{equation}
\int \frac{d^{D}p}{(2 \pi)^D}f(p)\rightarrow \frac{1}{\beta} \sum_{n} \int \frac{d^{D-1}p}{(2 \pi)^{D-1}}f(\frac{2n\pi}{\beta}, {\bf p}).
\label{sub}
\end{equation}
The thermal gap equation may be get from the 
zero temperature gap equation,  
\begin{equation}
M^2(\varphi)=m ^2+\frac{\lambda }{2}\varphi^2+ \frac{\lambda }{2} 
\int \frac{d^D k}{(2 \pi)^D} \frac{1}{k^2+M^2(\varphi)},
\label{18}
\end{equation}
after performing the replacement given by eq.(\ref{sub}). 
In the next section we use these result to analyse the thermal behavior 
of the squared mass and coupling constant. We remark that the previous 
analysis 
may also be done using a simpler but {\it ad-hoc} procedure \cite{drummond},
replacing the thermal mass obtained in the one-loop approximation by 
the Dyson-Schwinger equation. Nevertheless, the CJT 
formalism provides a more elegant way and and a consistent basis 
for this study, since the gap equation is 
derived from a stationary requirement.
Indeed, recently Campbell-Smith 
analysed the composite operator formalism in $(QED)_{3}$ and proved that it
reproduces the usual gap equation derived using the {\it ad-hoc} 
Dyson-Schwinger approach \cite{cam}.

\section{The thermal gap equation in the $\lambda\varphi^{4}_{D}$ model}\
Let us suppose that our system is in equilibrium with a thermal bath.
At the one-loop approximation the thermal mass and coupling 
constant for the $\lambda\varphi^{4}$ model in a D-dimensional 
Euclidean space have been obtained in a previous work \cite{adolfo} and 
are given by 
\begin{equation}
\ m^{2}(\beta)=m^{2}_{0}+
\frac{\lambda_{0}}{(2\pi)^{D/2}}
\sum^{\infty}_{n=1}\biggl(\frac{m_{0}}{\beta n}\biggr)^
{\frac{D}{2}-1}K_{\frac{D}{2}-1}(m_{0}n\beta)
\label{mass}
\end{equation}
and
\begin{equation}
\lambda(\beta)=
\lambda_{0}-\frac{3}{2}\frac{
\lambda^{2}_{0}}{(2\pi)^{D/2}}\sum^{\infty}_{n=1}
\biggl(\frac{m_{0}}{\beta n}\biggr)^
{\frac{D}{2}-2}K_{\frac{D}{2}-2}(m_{0}n\beta),
\label{coupling}
\end{equation}
where $K_{\nu}(z)$ is the modified Bessel function and $m_{0}^2$ and 
$\lambda_{0}$ are the zero temperature renormalized squared mass and
coupling constant respectively \cite{dif}. It is possible to improve the above results 
studying the gap equation
for the temperature dependent squared mass. 
One way to get this improvement is to take the 
$O(N)$ model, as it has been done by Dolan and Jackiw.
In this case the Feynman diagrams are classified according to 
its topology and each of these classes is associated to a given 
power of $\frac{1}{N}$. 
In the large $N$ limit, only some classes of diagrams, those 
associated to the smallest power of $\frac{1}{N}$ give the leading 
contributions. These contributions, which include those from all daisy and 
super-daisy diagrams may be summed up, resulting that 
in the large $N$ limit the results are exact. This approach was further 
developed by Weldon and also by Eboli et al. and more recently by
Drummond et al. \cite{drummond}.
We can obtain from the thermal gap equation resulting from the 
combination of eq.(\ref{sub}) and eq.(\ref{18}),  
an expression to replace the one-loop result above.
For simplicity we will first analyse the thermal corrections in 
the the disordered phase, which 
corresponds to absence of spontaneous symmetry breaking. 
In this case we adopt the 
notation $m^{2}(\beta)$ for the squared thermal mass and we suppress the 
quadratic term in the field.
We follow a procedure 
analogous to that Malbouisson and Svaiter have used in \cite{adolfo}. We use a mix 
between dimensional and zeta function analytical regularizations to 
evaluate formally the integral over the continuous 
momenta and the summation 
over the Matsubara frequencies. We get a sum of a polar (temperature 
independent) term plus a thermal analytic correction. The pole is suppressed 
by the renormalization procedure. Then after some technical manipulations, 
the gap equation may be rewritten in the form,
\begin{equation}
\ m^{2}(\beta)=m^{2}_{0}+
\frac{\lambda_{0}}{(2\pi)^{D/2}}
\sum^{\infty}_{n=1}\biggl(\frac{m(\beta)}{\beta n}\biggr)^
{\frac{D}{2}-1}K_{\frac{D}{2}-1}\left(m(\beta)n\beta\right).
\label{gap1}
\end{equation}
To solve the above  equation, and consequently to go beyond perturbation theory, let us take an integral representation of the Bessel function \cite{abra} given by
\begin{equation}
K_{\nu}(z)=
\frac{\pi^{\frac{1}{2}}}{\Gamma(\nu+\frac{1}{2})}
(\frac{1}{2}z)^{\nu}\int^{\infty}_{1}e^{-zt}(t^{2}-1)^{\nu-\frac{1}{2}}dt,
\label{def}
\end{equation}
which is valid for $Re(\nu)>-\frac{1}{2}$ and $|arg(z)|<\frac{\pi}{2}$. This integral representation attend for our 
purposes if we restrict ourselves to $D>1$. Substituting eq.(\ref{def}) in eq.(\ref{gap1}) and defining
\begin{equation}
F(D)=\frac{1}{2^{D-1}}\frac{1}
{\pi^{\frac{D-1}{2}}}\frac{1}{\Gamma(\frac{D-1}{2})},
\label{not}
\end{equation}
the gap equation becomes 
\begin{equation}
\ m^{2}(\beta)=m^{2}_{0}+
\lambda_{0} F(D)\left(m(\beta)\right)^{D-2}\int^{\infty}_{1}dt(t^{2}-1)^{\frac{D-3}{2}}
\frac{1}{e^{m(\beta)\beta t}-1}.
\label{gap2}
\end{equation}
Defining a new variable $\tau=m(\beta)\beta t$ it is easy to show that 
\begin{equation}
\ m^{2}(\beta)=m^{2}_{0}+
\lambda_{0} F(D) \left(m(\beta)\right)^{D-2}\int^{\infty}_{m(\beta)\beta}d\tau
\left((\frac{\tau}{m(\beta)\beta})^{2}-1\right)^{\frac{D-3}{2}}
\frac{1}{e^{\tau}-1}.
\label{gap3}
\end{equation}
When $D$ is odd, the power $\frac{D-3}{2}=p$ is an integer and the 
use of the Newton binomial theorem will give a very direct way for 
evaluating $m^{2}(\beta)$. When $D$ is even (the most interesting case) 
the expansion of $\left((\frac{\tau}{m(\beta)\beta})^{2}-1\right)^{\frac{D-3}{2}}$ 
yields a infinite power series, and  
the expression for the thermal squared mass becomes
\begin{equation}
\ m^{2}(\beta)=m^{2}_{0}+
\lambda_{0}\beta^{2-D}\sum^{\infty}_{k=0}f(D,k)\left(m(\beta)\beta\right)^{2k}
\int^{\infty}_{m(\beta)\beta}d\tau\frac{\tau^{D-3-2k}}{e^{\tau}-1},
\label{gap4}
\end{equation}
where 
\begin{equation}
f(D,k)=F(D)(-1)^{k}C^{k}_{\frac{D-3}{2}},
\label{efe}
\end{equation}
and
\begin{equation}
C^{0}_{p}=1,C^{1}_{p}=\frac{p}{1!},..C_{p}^{k}=
\frac{p(p-1)..(p-k+1)}{k!},
\label{comb}
\end{equation}
are a generalization of the binomial coeficients.
Note that for small values of k the integral that appear in 
eq.(\ref{gap4}) is a Debye integral of the type
\begin{equation}
I_{1}(x,n)=\int_{x}^{\infty}d\tau\frac{1}{e^{\tau}-1}\tau^{n}=
\sum_{q=1}^{\infty}e^{-qx}(\frac{x^{n}}{q}+\frac{nx^{n}}{q^{2}}+
...\frac{n!}{q^{n+1}}),
\label{debye}
\end{equation}
which is valid for $x>0$ and $n\geq 1$ \cite{abra}. 
For k satisfying $k>\frac{D-4}{2}$, which corresponds to $n<1$ in the 
preceeding equation, it is necessary to generalize the Debye integral 
(the case $n=0$ is trivial). Let us investigate the case $n<0$. This 
generalization has been done by
Svaiter and Svaiter \cite{svaiter} and the result reads,
\begin{equation}
I_{2}(x,n)=\int_{x}^{\infty}d\tau\frac{1}{e^{\tau}-1}\frac{1}{\tau^{n}}=
-\sum_{q=0,q\neq n}^{\infty}\frac{B_{q}}{q!}\frac{x^{q-n}}{q-n}-
\frac{1}{(n!)}B_{n}\ln x+\gamma_{\frac{n-1}{2}},
\label{gen}
\end{equation}
(for odd $n$),  $Re(x)>0$, $2\pi>|x|>0$ and $\gamma_{\frac{n-1}{2}}$ being 
a constant. The quantites $B_{n}$ are the Bernoulli numbers.
Note that this generalization can be 
used only for high-temperatures i.e. $m(\beta)\beta<2\pi$. Thus, 
in the high temperature regime, if we define 
\begin{equation}
I(x,D-3-2k)=
\left\{ 
\begin{array}{ll}
I_{1}(x,D-3-2k), \;\;\; for \;\;\; x>0, \;\;\;   k\leq \frac{D-4}{2} \\
I_{2}(x,D-3-2k), \;\;\; for \;\;\; 0<x<\pi, \;\;\;    k > \frac{D-4}{2},
\end{array}
\right.
\label{def1}
\end{equation}
we may write 
\begin{equation}
m^{2}(\beta)=m^{2}_{0}+
\lambda_{0}\beta^{2-D}
\sum_{k=0}^{\infty}f(D,k)\left(m(\beta)\beta\right)^{2k}
I\left(m(\beta)\beta,D-3-2k\right).
\label{gap5}
\end{equation}
The above equation gives a non-perturbative 
expression for the thermal squared mass in the high temperature regime 
in the case of even dimensional Euclidean space. In the odd dimensional 
case the summation in $k$ finishes at $\frac{D-3}{2}$.

For any dimension, it is possible to perform a numerical analysis of the  behavior of the renormalized squared mass for all
temperatures using eq.(\ref{gap1}).  
It is found that in both cases $D=3$, and $D=4$, the thermal squared 
mass appears as a positive monotonically increasing function of 
the temperature.
For a negative squared mass $m^{2}_{0}$, the model exhibits spontaneous 
breaking of the $O(N)$ symmetry to $O(N-1)$. Since the 
thermal correction to the squared mass is positive, the symmetry is restored 
at sufficiently high temperature. The critical temperature $\beta^{-1}_{c}$ 
is defined as the value of the temperature for which $M^{2}(\beta)$ 
vanishes. In the neighborhood of the critical temperature we can use 
without loss of generality, eq.(\ref{gap1}) instead of eq.(\ref{18}),
since the normalized expectation value of the field 
(the order parameter) vanishes.   
Using the limiting formula for small arguments of the 
Bessel function  it is not difficult to show that the critical temperature 
in a generic D-dimensional Euclidean space $(D>2)$ is given by:
\begin{equation}
(\beta_{c}^{-1})^{D-2}=-\frac{m^{2}_{0}}{\lambda g(D)},
\label{critical}
\end{equation}
where $g(D)=\frac{1}{4\pi^{\frac{D}{2}}}\Gamma(\frac{D}{2}-1)\zeta(D-2)$.
In $D=4$ the above result reproduces the known value of the critical 
temperature. For $D=3$ the zeta function in $g(D)$ has a pole and 
a renormalization procedure implies that the quantity $\beta^{D-2}$ is proportional to the 
the regular part of the analytic extension of the zeta function in the 
neighborhood of the pole. It is not difficult to show that in this case the 
critical temperature is given by $\frac{8\pi m^{2}}{\lambda_{0}}=\beta^{-1}_{c}ln(\mu^2\beta_{c}^{2})$. 
This result is in agreement with 
the estimate of Einhorn et al. \cite{jon}, which has been 
recently confirmed by numerical
simulation \cite{bimo}.
     
\section{The thermal coupling constant for the Vector  
$\lambda\varphi^{4}_{D}$ model}

In this section we  investigate the behavior of the renormalized thermal 
coupling constant of the vector N-component $\lambda\varphi^{4}_{D}$ model.
Again, withouth loss of generality, 
let us suppose that we are in the symmetric phase i.e $m^{2}_{0}>0$. 
To go beyond perturbation theory, we take the leading order in
$\frac{1}{N}$, in which case we know that the contributions come only from 
some classes of diagrams (the chains of elementary four-point bubbles)  
and that it is it is possible to perform
summations over them. Proceeding in that way,
we get for the thermal renormalized coupling constant an expression 
of the form,   
\begin{equation}
\lambda(\beta)=
\frac{\lambda_{0}}{1-\lambda_{0}L\left(m^{2}(\beta),\beta\right)},
\label{cou1}
\end{equation}
where
\begin{equation}
L\left(m^{2}(\beta),\beta\right)=-\frac{3}{2}
\frac{1}{(2\pi)^{D/2}}\sum^{\infty}_{n=1}
\biggl(\frac{m(\beta)}{\beta n}\biggr)^
{\frac{D}{2}-2}K_{\frac{D}{2}-2}\left(m(\beta)n\beta\right),
\label{cou2}
\end{equation}
and $\lambda_{0}$ is the zero-temperature renormalized coupling constant.
For simplicity we have suppresed the factor $\frac{1}{N}$ everywhere. 
Incidentally, before 
going into more details of the non-perturbative finite temperature case, 
some comments are in order. 
We remark that at zero temperature and $D=4$ it is known that 
the theory has tachyons, which are related to the occurence of a Landau 
pole in $\lambda(m^{2})$. The conventional way to circunvect this 
problem is to interpret the model as an effective theory (introducing a ultraviolet cut-off) which restricts the energy to a 
region far below the tachyon mass. 
This is a necessary procedure, to take into account the widelly 
spread conjecture that the $\lambda\varphi^{4}$ 
model has a trivial continuum 
limit at $D=4$ and is meaningful only as an effective theory. 
Coming back to finite temperature,
firstly let us investigate the thermal behavior of the coupling constant  
in a Euclidean space satisfying $D>3$.
Proceeding in a manner analogous as we have done in the previous
section, we use again an integral representation of the 
Bessel function given by eq.(\ref{def}), which leads to the result,
\begin{equation}
L(m^{2}\left(\beta),\beta\right)=
G(D)\left(m(\beta)\right)^{D-4}\int^{\infty}_{1}dt(t^{2}-1)^{\frac{D-5}{2}}
\frac{1}{e^{m(\beta)\beta t}-1}
\label{cou3}
\end{equation}
where 
\begin{equation}
G(D)=-\frac{3}{2}
\frac{1}{(2\sqrt{\pi})^{D-1}}\frac{1}{\Gamma(\frac{D-3}{2})}.
\label{menos}
\end{equation}
Note that there are no poles in the Gamma function and $G(D)$ never 
vanishes. Defining 
\begin{equation}
g(D,k)=G(D)(-1)^{k}C^{k}_{\frac{D-5}{2}}
\label{mais}
\end{equation}
it is not dificult to 
show that 
\begin{equation}
L\left(m^{2}(\beta),\beta\right)=
\beta^{-D}\sum^{\infty}_{k=0}g(D,k)\left(m(\beta)\beta\right)^{2k+2}
\int^{\infty}_{m(\beta)\beta}d\tau\frac{\tau^{D-5-2k}}{e^{\tau}-1}.
\label{cou4}
\end{equation}
A straigthforward calculation gives us the following 
expression
\begin{equation}
L\left(m^{2}(\beta),\beta\right)=
\beta^{-D}
\sum_{k=0}^{\infty}g(D,k)\left(m(\beta)\beta\right)^{2k+2}
I\left(m(\beta)\beta,D-5-2k\right),
\label{fim}
\end{equation}
where again $I(x,n)$ is defined in eq.(\ref{def1}).
Then, substituting eq.(\ref{fim}) in the eq.(\ref{cou1}) we get the 
high temperature thermal coupling constant for $D>3$. In the case
$D \leq 3$, the integral representation of the 
Bessel function gived by 
eq.(\ref{def}) can not be used. Consequently, we take 
another integral representation of the Bessel function i.e.
\begin{equation}
K_{\nu}(z)=\frac{1}{2}\left(\frac{z}{2}\right)^{\nu}
\int_{0}^{\infty}dt\; e^{-t-\frac{z^2}{4t}}\;t^{-(\nu+1)}
\label{def2}
\end{equation}  
which is valid for $|arg(z)|<\frac{\pi}{2}$ and $Re(z^{2})>0$. 
A straighforward calculation gives 
\begin{equation}
L\left(m^{2}(\beta),\beta\right)=Q(D)m(\beta)^{D-4}\int_{0}^{\infty}
dt\; e^{-t}\; t^{-\frac{D}{2}+1}\left(\Theta_3(\pi,e^{-\frac{m^{2}\beta^{2}}
{4t}})-1\right),
\label{c3}
\end{equation}
where the theta function $\Theta_{3}(z,q)$ is defined by \cite{abra}:
\begin{equation}
\Theta_{3}(z,q)=1+2\sum_{n=1}^{\infty}q^{n^{2}}\cos(2nz).
\label{c4}
\end{equation}
and $Q(D)=-\frac{3}{2}\frac{1}{(2\sqrt{\pi})^D}$.
Substituting $L\left(m^{2}(\beta),\beta\right)$ in eq.(\ref{cou1}) we have  
a closed expression for the thermal renormalized coupling constant in the 
case $D\leq 3$. It is interesting to note that a long time ago 
Braden \cite{braden} has employed also the theta function $\Theta_{3}(z,q)$, to investigate the finite temperature $\lambda\varphi^{4}$ model.
We would like to stress that
the behavior of the thermal renormalized coupling constant 
is quite different from the monotonically increasing in temperature behavior  
obtained for the squared mass. Indeed, the 
kind of the thermal behavior of the coupling constant depends on the 
Euclidean dimension.
For $D=3$ the coupling constant 
(as a function of the temperature) decrease until 
some minimum value and then start to increase. 
This thermal behavior of the coupling constant is ploted in fig.(1).
For $D=4$ the 
thermal renormalized coupling constant tends to a constant value in the 
high temperature limit. See fig.(2).
Our result is consistent with the work of Fendley \cite{fen}.

It must be noticed that the thermal behaviour of the coupling
constant is very sensitive to the thermal behaviour of the mass.
As an ilustration of this fact we exhibit in fig.(3) the general aspect 
of the coupling constant as function of the temperature for
the same model we have treated here, but subjected to Wick ordering 
\cite{de}. In this case all tadpoles are suppressed and the 
thermal behaviour of the coupling constant does not depend at all on 
the mass thermal behaviour. We see that in this situation the coupling 
constant is a monotonic decreasing function of the temperature. The 
absence of Wick ordering deeply changes this behaviour. We show
in fig.(4) for $D=3$ in the same scale the plots for $\lambda(T)$ with and
without Wick ordering, respectivelly the lower and the upper curves.
In the region of temperatures where $\lambda_{W}(T)$ goes
practically to zero, $\lambda(T)$ is practically constant at a value
slightly lower than the common zero-temperature coupling constant
$\lambda_{0}$. The growth of $\lambda(T)$ with the temperature
presented in fig.(1) is in a much smaller scale for $\lambda(T)$
than in fig.(4). In fact this growth is "microscopic" in a scale 
where the Wick ordered coupling constant $\lambda_{W}(T)$ presents
asymptotic thermal freedom.

\section{Conclusions}

We have done in this paper an analysis of the vector 
$\lambda\varphi^{4}$ model 
in a flat D-dimensional Euclidean space  in equilibrium with a thermal bath. 
The form of the thermal corrections to the mass 
and coupling constant have been discussed using resummation methods. 
We have formulated the problem in a general framework, but 
our results are at leading order in the $\frac{1}{N}$ expansion.    
We have chosen this way of working, in order to get answers as much as 
possible of a non-perturbative character. In what concerns the thermal 
mass behavior,
we have shown that the thermal renormalized squared mass is a monotonic 
increasing function of the temperature for any Euclidean dimension. 
We have been able to obtain a general formula for the critical temperature 
of the second order phase transition valid for any Euclidean dimension 
$D>2$, provided the necessary renormalization procedure is done 
to circunvect singularities of the zeta function. The values obtained 
for the critical temperatures in $D=3$ and $D=4$ agree with previous results.
     
The 
behavior of the thermal coupling constant depends on the dimensionality of
the Euclidean space. In $D=3$ the renormalized coupling 
constant decreases until some positive 
minimum value and then starts to increase slightly as a function 
of the temperature. See fig.(1). This 
result seems to indicate that at a non-perturbative level (for large N) 
in the framework of the vector N-component model, the 
answer to the question raised in Ref.(\cite{adolfo}) is negative: 
there is no first order phase transition induced by the 
thermal coupling constant in $D=3$.
In $D=4$ the thermal renormalized coupling constant in the high temperature limit tends to a constant 
value, $(\lambda_{0}-(3\frac{\sqrt{6}}{8\pi})\lambda_{0}^{\frac{3}{2}}$,
which coincides exactly with the result obtained by Fendley \cite{fen}).

A natural extension of this work should be to go beyond the $\frac{1}{N}$ leading order results using renormalization group methods. 
Another possible direction is to introduce an abelian gauge field 
coupled to the 
N-component scalar field. In $D=3$ a topological Chern-Simons term 
may be added, 
and also a $\varphi^{6}$ term. In this case we have shown  in a previous work 
using perturbative and semi-classical techniques that the 
topological mass makes appear a richer phase structure introducing the 
possibility of first or second order phase transitions depending on the 
value of the topological mass \cite{fla}. The use of resummation methods to
investigate the thermal behavior of the physical quantities could 
generalize these previous results to the N-component vector model. 
These will be subjects of future investigation.

\section{Acknowlegements}

We would like to thank M.B.Silva-Neto and C.de Calan by 
fruitfull discussions.
This paper was supported by Conselho Nacional de 
Desenvolvimento Cientifico e Tecnologico do Brazil (CNPq).

\newpage

\begin{figure}[th]

\centerline{\epsfxsize=4in\epsfysize=3in\epsffile{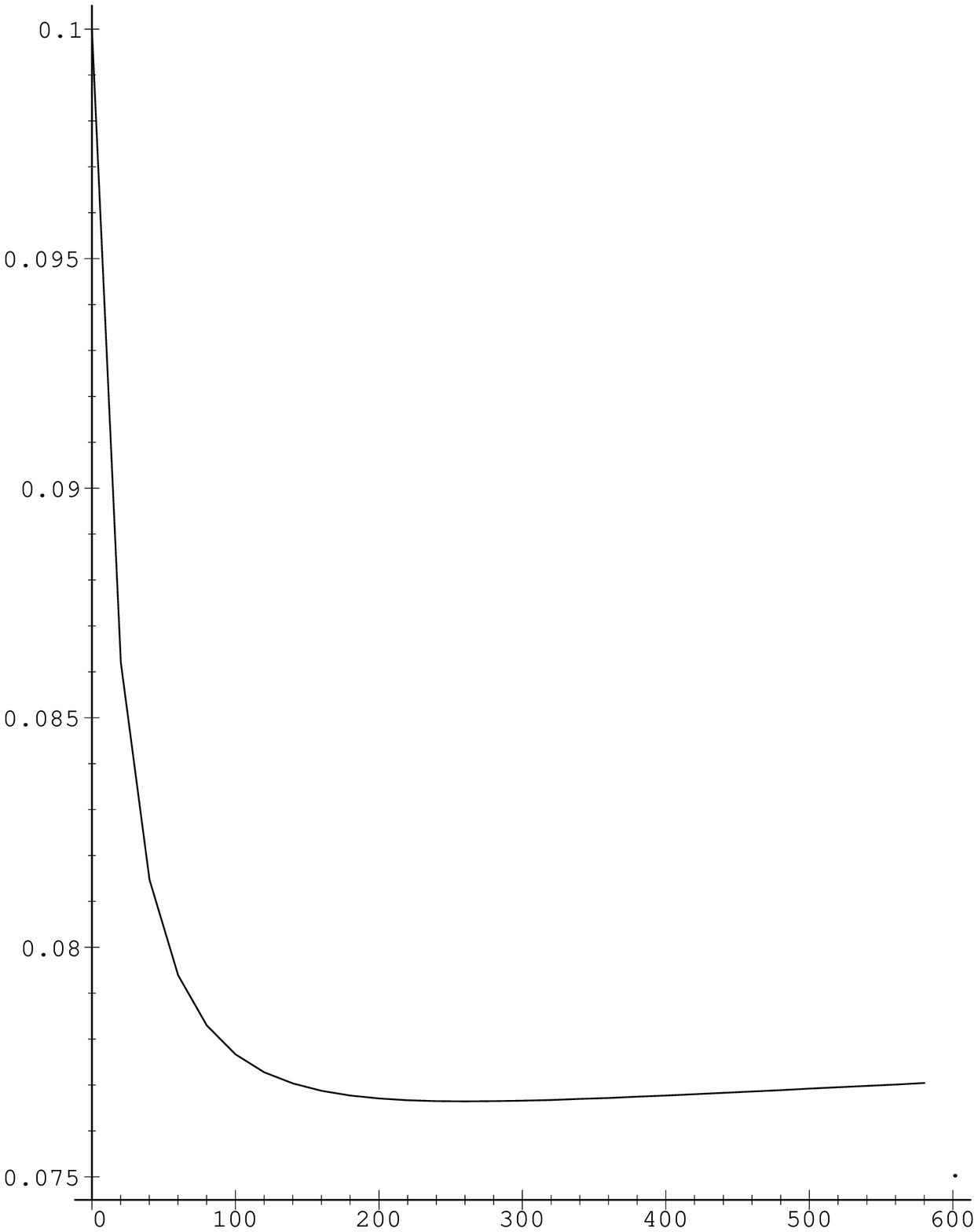}} 
\caption[region]
{Coupling constant thermal behavior obtained from eqs. (\ref{gap1}),(\ref{cou1}) and (\ref{cou2})  in dimension $D=3$ .}
%\end{figure}

%\begin{figure}[h]
\centerline{\epsfxsize=4in\epsfysize=3in\epsffile{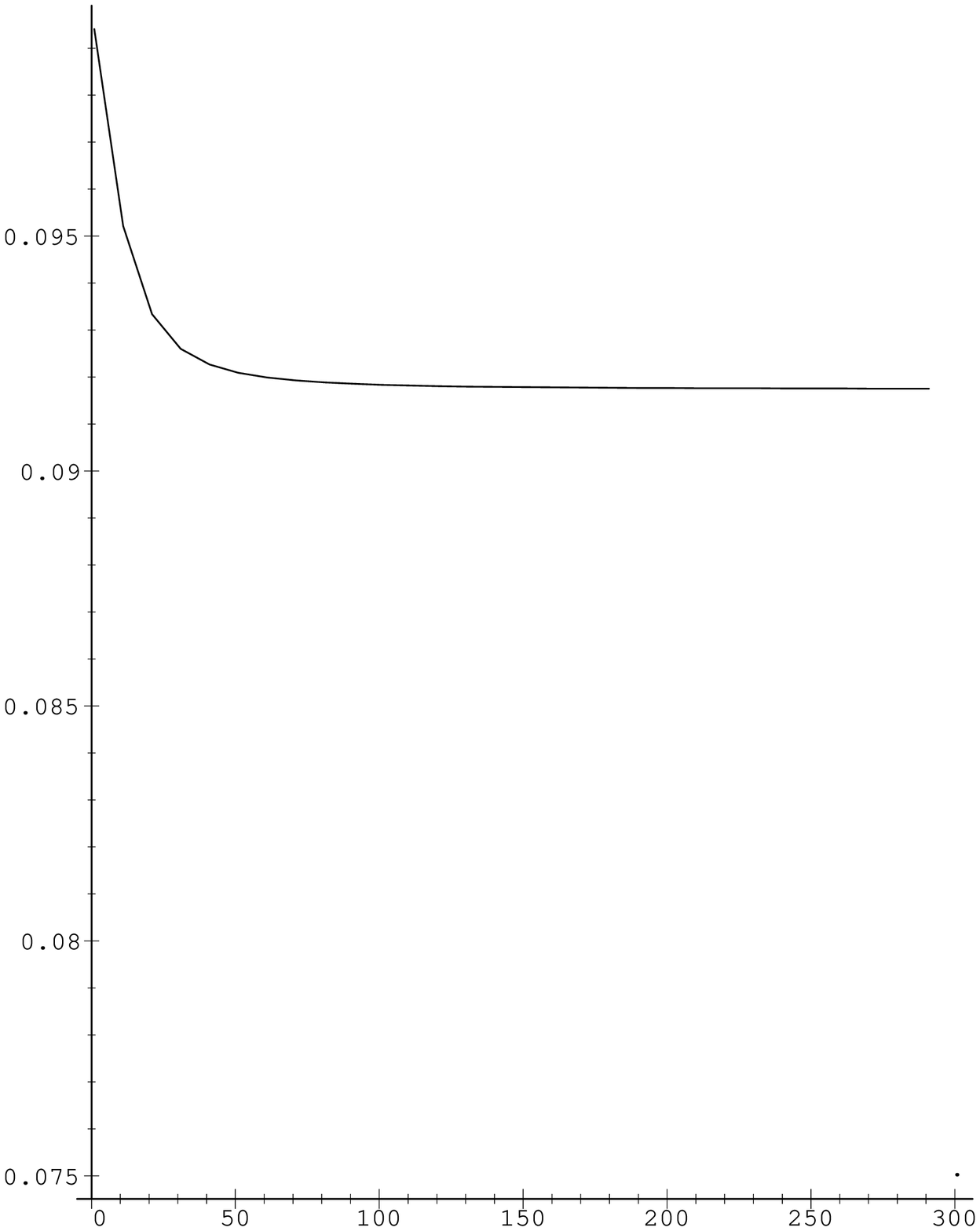}} 
\caption[region]
{Coupling constant thermal behavior obtained from eqs. (\ref{gap1}),(\ref{cou1}) and (\ref{cou2})  in dimension $D=4$.}
\begin{picture}(10,10)
%\put(0,0){0}
\put(160,540){{\small $\lambda (T)$}}

\put(160,260){{\small $\lambda (T)$}}

\put(350,95){{\small $T$}}
\put(350,375){{\small $T$}}

\end{picture}

\end{figure}

\begin{figure}[th]
%\begin{figure}[h]
\centerline{\epsfxsize=4in\epsfysize=3in\epsffile{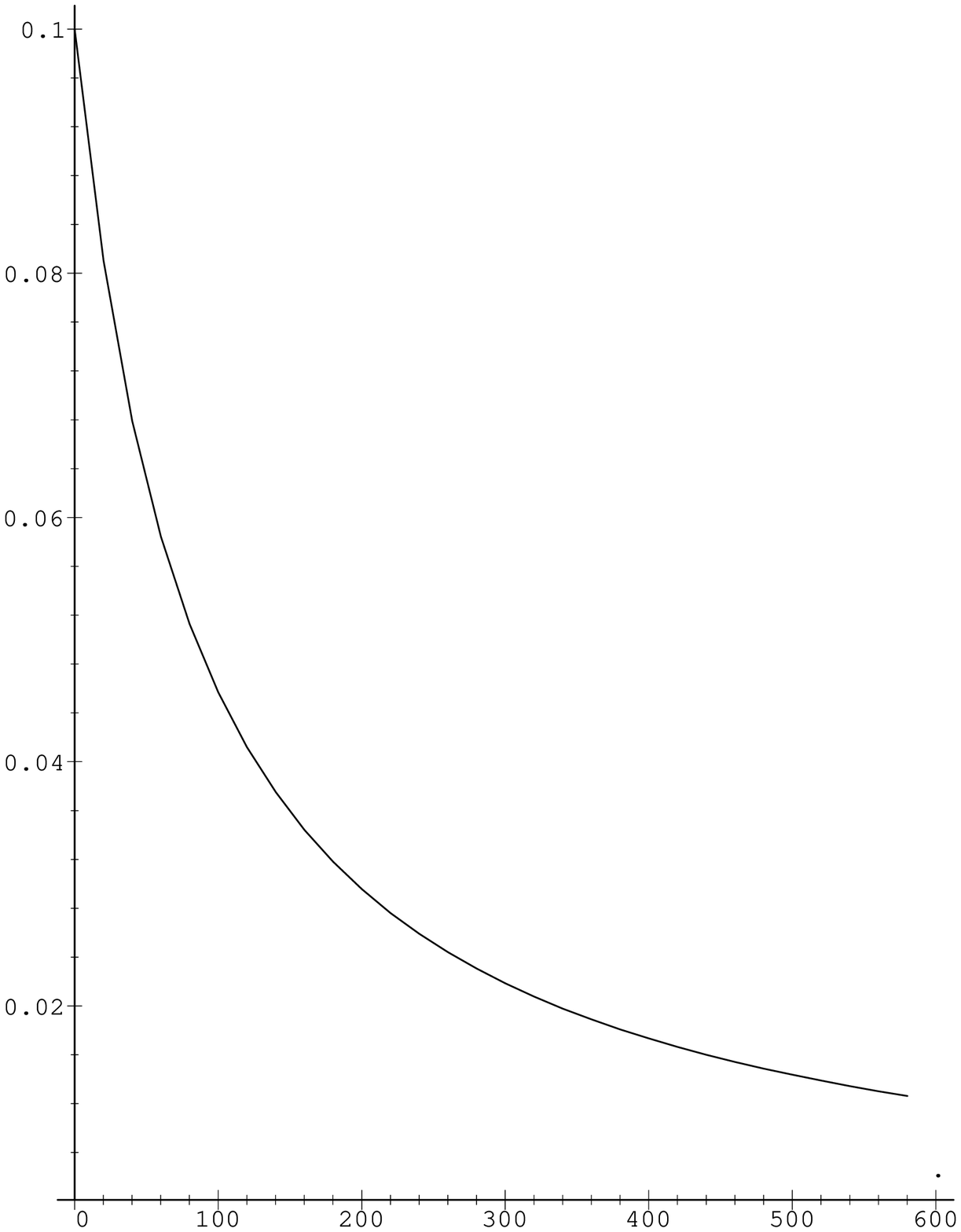}} 
\caption[region]
{General aspect of coupling constant thermal behavior 
obtained from eqs. (\ref{gap1}),(\ref{cou1}) and (\ref{cou2}) for the Wick ordered model.}
%\end{figure}

%\begin{figure}[h]
\centerline{\epsfxsize=4in\epsfysize=3in\epsffile{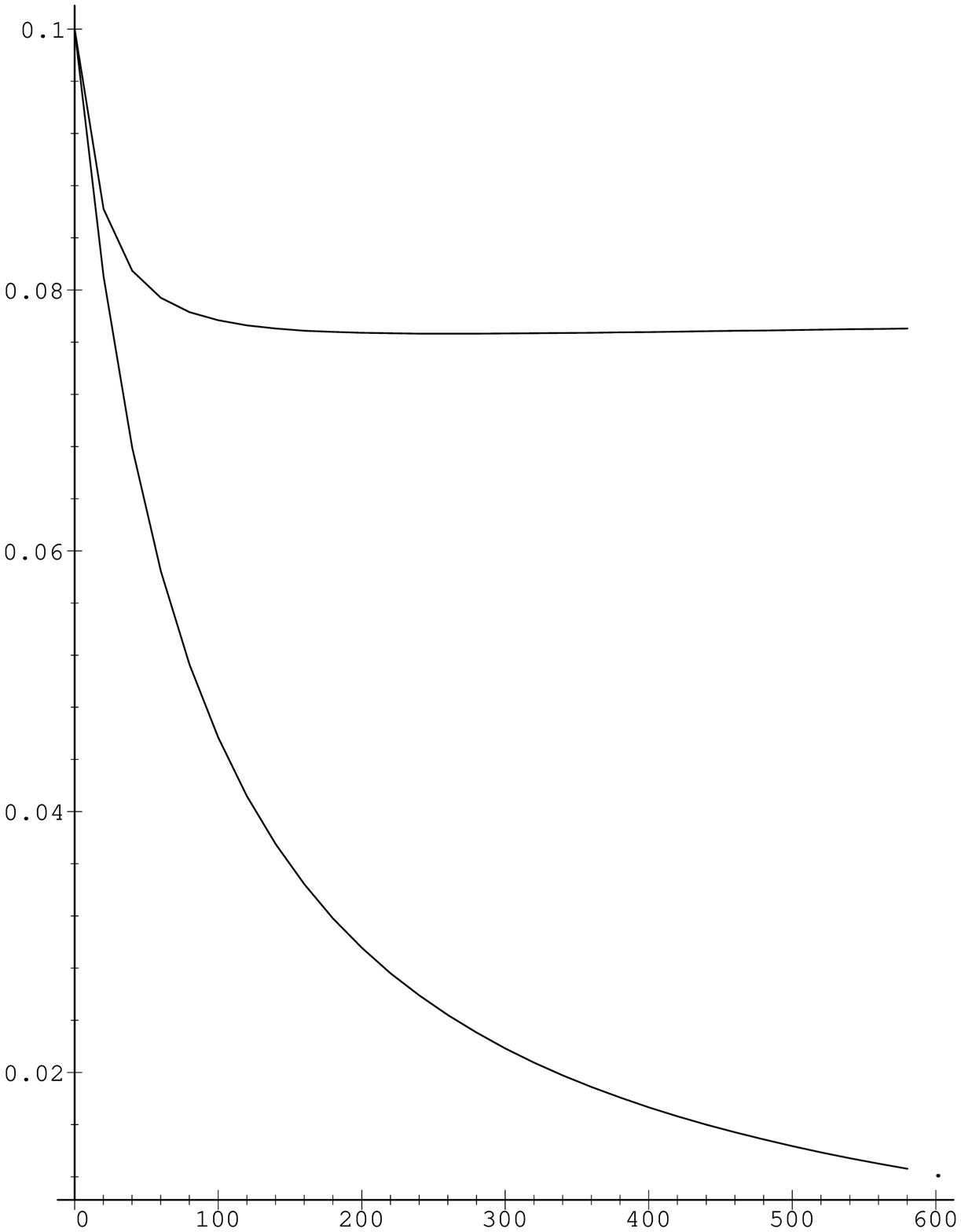}} 
\caption[region]
{Compared thermal behaviors of the coupling constant for 
the Wick-ordered model and non Wick-ordered model.}
\begin{picture}(10,10)
%\put(0,0){0}
\put(160,540){{\small $\lambda_W (T)$}}

\put(200,230){{\small $\lambda (T)$}}
\put(200,160){{\small $\lambda_W (T)$}}
\put(350,95){{\small $T$}}
\put(350,375){{\small $T$}}

\end{picture}

\end{figure}

\end{document}